\documentstyle[12pt,epsf]{article}   
\newcommand {\trd}{\mbox {$\frac{1}{3}$}}
\newcommand {\ttd}{\mbox {$\frac{2}{3}$}}

\newcommand {\soh}{\mbox {${\rm s}_{\frac{1}{2}}$}}
\newcommand {\poh}{\mbox {${\rm p}_{\frac{1}{2}}$}}
\newcommand {\pth}{\mbox {${\rm p}_{\frac{3}{2}}$}}
\newcommand {\dth}{\mbox {${\rm d}_{\frac{3}{2}}$}}
\newcommand {\dfh}{\mbox {${\rm d}_{\frac{5}{2}}$}}

\newcommand {\fsh}{\mbox {${\rm f}_{\frac{7}{2}}$}}
\def\eqa{\begin{eqnarray}}
\def\qea{\end{eqnarray}}
\def\eqref#1{eq.~(\ref{eq:#1})}

\newcommand {\eq}{\begin{equation}}
\newcommand {\qe}{\end{equation}}

\newcommand {\ajt}{\mbox {$\tilde{a}_{j}$}}

\newcommand {\bjd}{\mbox {$b_{j}^{\dagger}$}}

\newcommand {\vk}{\mbox{\boldmath $k$}}
\newcommand {\vkp}{\mbox {\boldmath $k'$}}
\newcommand {\vr}{\mbox {\boldmath $r$}}

\newcommand {\ost}{\mbox {$^{16}{\rm O}$}}
\newcommand {\oet}{\mbox {$^{18}{\rm O}$}}
\newcommand {\nest}{\mbox {$^{16}{\rm Ne}$}}
\newcommand {\neet}{\mbox {$^{18}{\rm Ne}$}}
\newcommand {\caft}{\mbox {$^{40}{\rm Ca}$}}
\newcommand {\caftt}{\mbox {$^{42}{\rm Ca}$}}
\newcommand {\cate}{\mbox {$^{38}{\rm Ca}$}}
\newcommand {\tift}{\mbox {$^{40}{\rm Ti}$}}
\newcommand {\cft}{\mbox {$^{14}{\rm C}$}}

\newcommand {\ostne}{\mbox {$^{16}{\rm O}(\pi^{+},\pi^{-})^{16}{\rm Ne}\ $}}
\newcommand {\oetne}{\mbox {$^{18}{\rm O}(\pi^{+},\pi^{-})^{18}{\rm Ne}\ $}}
\newcommand {\caftti}{\mbox {$^{40}{\rm Ca}(\pi^{+},\pi^{-})^{40}{\rm Ti}\ $}}
\begin{document}

\begin{center} 
{\bf Double Charge Exchange And Configuration Mixing}\\

\vspace{0.4cm} 
 H.C. Wu 

Instituto de Fisica, Universidad de Antioquia,\\
AA 1226 Medellin, Colombia\\

\vspace{0.4cm} 
W.R. Gibbs

Physics Department, New Mexico State University, \\
Las Cruces, NM 88003, USA

\vspace{0.4cm} 
 
\today

\abstract {The energy dependence of forward pion double charge exchange
reactions on light nuclei is studied for both the Ground State transition
and the Double-Isobaric-Analog-State transitions. A common characteristic
of these double reactions is a resonance-like peak around 50 MeV pion lab
energy.  This peak arises naturally in a two-step process in the
conventional pion-nucleon system with proper handling of nuclear
structure and pion distortion.  A comparison among the results of
different nuclear structure models demonstrates the effects of
configuration mixing. The angular distribution is used to fix the single
particle wave function.}

 PACS: 25.80Gn, 24.10Ht,21.60Fw
 
\end{center} 
 
\section{Introduction}

The energy dependence of the forward angle cross section for pion-nucleus
double charge exchange (DCX) reactions has been shown to contain important
information on the fundamental pion-nucleon charge exchange process and
has been investigated extensively both experimentally
\cite{dra00,fol97,pat98} and theoretically \cite{bil93,nus98}. See Ref.
\cite{mikkel} for a review of the field in 1993.

The systematic data on the energy dependence of the DCX reaction
consistently show a peak at low energy, which exists for both the double
isobaric analog state (DIAS) and the ground state (GS) transitions.  
There have been controversial theoretical explanations of this peak. While
some authors\cite{bil93} have claimed that the peak is evidence for the
existence of a dibaryon, it has been shown \cite{nus98} that it can be
understood based on a two-step sequential process which involves solely
conventional degrees of freedom of pions and nucleons. In Ref.
\cite{nus98}, a study on $^{42-48}$Ca and $\cft$ showed that the
appearance of this peak at low energy is naturally explained by the
distortion of the pion scattering wave without invoking the existence of a
dibaryon.

The purpose of the present study is to investigate the question of what
information about nuclear structure can be obtained from the energy
dependence of the forward DCX reaction, more specifically, what is the
effect of configuration mixing on the energy dependence. In the present
work we demonstrate the relevance of configuration mixing through a
comparison of different nuclear models. In particular we use the shell
model and models of extreme L-S and extreme j-j coupling.

Early work on $\ost$ indicated the importance of nuclear structure in the
DCX reaction\cite{lee77}.  In that case the calculation was based on the
plane wave approximation for the pion wave function. One understands now
that the distortion of the pion wave (by an optical potential) plays a
very important role in the DCX reaction.  For a quantitative understanding
of the effect of nuclear structure on DCX reactions a suitable optical
model is indispensable.

The present work uses the theoretical framework of the sequential
process, as was done in \cite{nus98}. For the pion wave distortion we
use the optical model\cite{wuh00} that originated from
Ref.\cite{gar81,kau83}, as extended to the multi-j case.

The article is organized as follows: The second section discusses the GS
transition \ostne \ and \caftti. The third section studies the DIAS
transition \oetne. The fourth section contains conclusions and discussion.

\section{Ground State Transitions: $\ost$ and $\caft$}

\subsection{Nuclear structure}

For the GS transitions \ostne and \caftti, the initial states can be
assumed to be closed shell states so that one has to deal with
configuration mixing only in the final states. The nuclear structure of
the two final state nuclei has been investigated by several authors
\cite{zuk68,law76,coh65,coh67}. Here we apply those results and compare
them to the results derived from the models of extreme L-S and j-j
coupling.

The SU(4) $\otimes$ SU(3) model is a limiting case of L-S coupling, where
SU(4) stands for the Wigner supermultiplet SU(4) symmetry \cite{wig37}
which assumes the degeneracy of all four spin-isospin states of the
nucleon, and SU(3) stands for the Elliott model\cite{ell58} which deals
with the (orbital) rotational symmetry.  A combination of these two
symmetries, i.e. the SU(4) $\otimes$ SU(3)  model (denoted as SU(3)
hereafter), works well in the p-shell and fairly well in the ds
shell \cite{par78}.

Another limiting case of nuclear structure is the single particle shell
model (SPSM) which puts nucleons into j-orbits solely according to the
single particle energy, neglecting the two-body couplings.  This is an
extreme case of j-j coupling.  The shell model lies between these two
extremes as far as the single particle occupancy in orbits is concerned
\cite{coh67}.  In general the shell model is more realistic than the
SU(3) and SPSM models.

For the DCX reaction on $\ost$ the initial and final states are the
ground state of $\ost$ and $\nest$, respectively. The simplest ground
state of $\ost$ is a closed shell. For a better description of the
excited states of $\ost$, a mixing of ds orbits was introduced
\cite{zuk68}. However, this ds mixing will contribute mainly to the
excited states, whereas our interest is only in the ground state of
$\ost$. Therefore, for the study of the DCX reaction on $\ost$, we
restrict ourselves to the closed shell state.

For $\nest$, we use the weak coupling approximation \cite{law80}, i.e.
the ground state of $\nest$ is taken as consisting of two protons in the
ds shell plus two neutron holes in the p shell. We make the further
simplification that only the angular momentum zero pairs of the proton
and neutron sector are considered.

In Ref. \cite{lee77} a truncated \poh-\dfh-\soh\ space was used.
Due to the large difference in the shapes of the energy dependence 
(see Section 3) of forward DCX amplitude among different configurations 
we pursue a description in the full p- and ds-shell.

We begin with the SU(3) scheme where the ground state of the proton sector 
can be written as,
\eq
|{\rm gs},p> _{\rm SU3} = |[11]01;(40)0;0>,
\qe
in which the symbol at the right hand side, $|[f]ST;(\lambda \mu)L;J>$,
contains the following quantum numbers: [f] and $(\lambda \mu)$ are the
SU(4) and SU(3) irreps, respectively; $ST$ represent the spin and isospin
of the two nucleon state, $L$ the orbital angular momentum, and $J$ the
total angular momentum.  The ground (hole) state of the
neutron sector is given by
\eq
|{\rm gs},n>_{\rm SU3}= |[11]01;(02)0;0>.
\qe
The above SU(3) representation can be transformed into the  j-j coupling 
formalism as follows,
\eqa
|{\rm gs},p>_{\rm SU3} &=& \frac{\sqrt{5}}{3} |(\soh)^{2}; 0>
+ \frac{2}{3} \sqrt{\frac{2}{5}} |(\dth )^{2}; 0>
+ \frac{2}{3} \sqrt{\frac{3}{5}} |(\dfh)^{2}; 0>
 \nonumber \\
|{\rm gs},n>_{\rm SU3} &=&  \sqrt{\trd}|(\poh )^{-2}; 0> + 
\sqrt{\ttd} |(\pth)^{-2}; 0>,
\qea
where, for example, $|(\soh)^{2}; 0>$ \ and $|(\pth)^{-2}; 0>$ \ stand for
a state with two particles in \soh \ orbit and a state with two holes in
the \pth \ orbit, respectively; and both states have angular momentum
zero.

The shell model describes a more ``realistic'' case, and normally
contains a mixture in the SU(3) irreps. We take the occupancy of single
particle orbits from previous studies\cite{zuk68,law76,coh65,coh67}.  The
probabilities of the hole state in the \poh and \pth\ orbits are 0.83 and
0.17, respectively.  Therefore, one can write the shell model hole
state as

\eq
|{\rm gs}, n>_{\rm SM} =  0.91 |(\poh)^{-2}; 0 > + 0.41 |(\pth)^{-2}; 0 >,
\qe
where the phase is determined by requiring that the $L=0$ term be
dominant.  In a similar manner the proton sector shell model wave function
is determined as

\eq
|{\rm gs}, p>_{\rm SM} =  0.32 |(\soh )^2; 0> + 0.32 |(\dth)^2; 0> +
0.89 |(\dfh)^2; 0>.
\qe

Another limiting case of the nuclear structure is the single particle
shell model, in which the two protons occupy the \dfh \ and the two
neutron holes occupy the \poh \ orbit.

The amplitudes of proton pairs in ds-shell orbits and the amplitudes of 
neutron hole pairs in
p-shell orbits are listed in Table 1, where the last three rows give the
amplitudes for the SU(3) model, the shell model (SM) and the single
particle shell model (SPSM), respectively.

\begin{table}[ht]
\vspace*{-12pt}
\begin{center}
\caption[]{
Single particle distribution over orbits}
\vspace*{14pt}
\begin{tabular}{|c|c|c|c|c|c|c|c|c|}
\hline
 &\multicolumn{5}{|c|}{$\nest$} & \multicolumn{3}{|c|}{$\tift$} \\ \hline
 &\multicolumn{3}{|c|} {proton} &\multicolumn{2}{|c|} {neutron(hole)}
 & \multicolumn{3}{|c|} {neutron(hole)} \\ \hline
 &\soh & \dth & \dfh & \poh & \pth & \soh & \dth & \dfh \\
\hline
 SU(3) &0.746 &0.421 &0.517 &0.577 &0.817 &0.746 &0.421 &0.517 \\ \hline
 SM  &0.319 &0.319 &0.892 &0.910 &0.415 & 0.224    &0.975 &0 \\ \hline
 SPSM  & 0    & 0    & 1.0  & 1.0  & 0    & 0    & 1.0    & 0\\ \hline
 \end{tabular}
\label{Ta:multicol+cline}
\end{center}
\end{table}

\begin{figure}[ht]
\epsfysize=110mm
\hspace*{2.0cm}
\epsffile{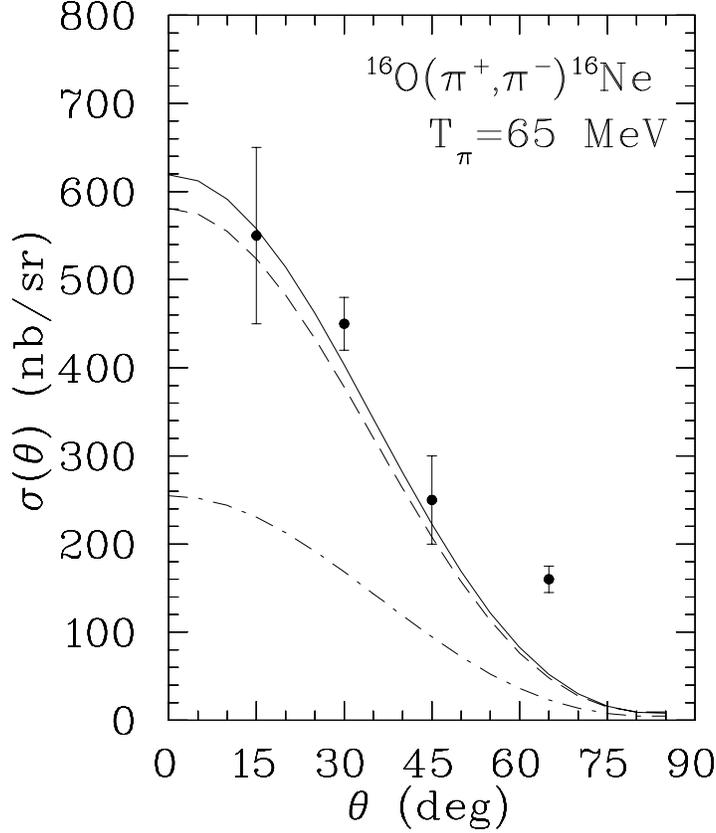}
\caption{Angular distribution of the reaction \ostne at a pion 
laboratory energy of 65 MeV. The solid line represents the result of 
shell model, the dashed line gives the result of the SU(3) model and the 
dashed-dotted line the result of single particle shell model.}
\label{f1}
\end{figure}

The ground state of $\caft$ is taken as a closed shell state, whereas the
ground state of $\tift$ is considered as consisting of two proton
particles in the \fsh \ shell and two neutron holes in the ds shell.  
The proton sector of $\tift$ is restricted to contain only one orbit,
\fsh, thus one only needs to deal with the configuration mixing in the
neutron sector.  The structure of the neutron sector is approximated by
that of the nucleus $\cate$. In the SU(3) scheme the ground state of the
neutron sector is
\eq 
|{\rm gs},n>_{\rm SU3} = |[11]01; (04)0; 0>,
\qe
where the notation is the same as for $\nest$. 

For the shell model we carry out a schematic calculation similar to that
for the structure of $\oet$ \cite{law80}, where the surface delta function
interaction is assumed. The ground state of $^{39}$Ca, $J^{\pi}
=\frac{3}{2}^{+}$ is the single hole state of $(\dth)^{-1}$, whereas the
excited state $J^{\pi} =\frac{1}{2}^{+}$ can be taken as the single hole
excited state $(\soh)^{-1}$. There being no excitation of
$\frac{5}{2}^{+}$ at low energy, we restrict ourselves to the truncated
hole space $(\dth-\soh)^{-1}$ . By imposing the requirement on the
equality of the calculated and experimental binding energies\cite{law80},
the ground state of $^{38}$Ca can be calculated as,

\eq
|{\rm gs}, n>_{\rm SM} =0.975|(\dth )^{-2}; 0> + 0.222 |(\soh )^{-2}; 0>,
\qe
where the proton closed shell is neglected. 

In the single particle shell model, the two neutron holes of $\tift$ are
in the \dth \ orbit. The amplitudes of neutron hole pairs 
 for $\tift$ are also listed in Table 1. The proton sector is not
given since there is only one orbit, \fsh.

It can be seen from the results of both $\nest$ and $\tift$\ that, as far
as the amplitude of particle (hole) pairs in different orbits is concerned, 
the shell model provides an intermediate case between the SU(3) model 
and the single particle shell model, 
as was shown in a previous study \cite{coh67}.

\subsection{Angular Distribution}

The calculation of the DCX reaction follows a procedure developed in a
previous paper \cite{wuh00}, which extends the work in \cite{aue88} (AGGK) to
a multi-j case.  Here we give the outline of this method. The DCX operator
(in coordinate space) is expressed as
\eq
 F_{12}(\vk, \vkp)=[{\cal F}_0 (\vr_1 ,\vr_2 )
     + {\cal F}_1 (\vr_1, \vr_2 ) {\bf \sigma_1 } \cdot {\bf e_1 }
  {\bf \sigma_2 } \cdot {\bf e_2 } ]
    T_{-} (1) T_{-} (2),
\qe
where ${\cal F}_0$ and  ${\cal F}_1$ are related to non-spin-flip
(NSF) and double-spin-flip (DSF) processes, respectively, and they are
calculated numerically treating pion wave propagation, distortion
and charge exchanges.  The neglect of the single-spin-flip term is for 
simplicity and is thought to be a reasonable approximation \cite{wuh00}.
In the multi-j case it is more convenient to express the DCX operator
in pair form:
\eq
F= \sum_{J,\{ j_1 j_2 \}, \{ j_3 j_4 \}}
 G_{J,\{ j_1 j_2 \}, \{ j_3 j_4 \} } (\bjd_1  \bjd_2 )^J
 \cdot (\ajt_3 \ajt_4 )^J ,
\qe
where $\{ j_1 j_2 \}$ means $j_1 \ge j_2$.
The two-body DCX matrix elements $ G_{J,\{ j_1 j_2 \}, \{ j_3 j_4 \} }$ 
consist of NSF and DSF terms for which the formula  can be found in Ref. 
\cite{wuh00}.

In multipole form the DCX operator looks the same as in the AGGK paper,
\eq
F=\Omega \sum_{L,\{ j_1 j_2 \}, \{ j_3 j_4 \}}
F_{L \{ j_1 j_2 \}, \{ j_3 j_4 \}} (\vk ,\vkp )
(\bjd_1  \ajt_3 )^L \cdot (\bjd_2 \ajt_4 )^L /(2L+1),
\qe
however, the actual expression of the matrix elements $F_L$ is more 
complex and can be derived from the pair-form matrix elements 
$G_{J,\{j_1 j_2\} \{ j_3 j_4 \}} $.

In calculating the angular distribution, the parameters of the optical
potential are taken from Ref. \cite{nus98} which presents the result of
fitting the elastic scattering. Another important ingredient in the
calculation is the single particle wave function, especially the wave
function of the orbits which are involved in the DCX reaction.  The single
particle wave functions are calculated by using a Woods-Saxon type of
potential, the parameters being so adjusted that both the separation
energy of the proton and neutron and the rms proton radius match the 
experimental values.  However, this is still not enough to determine the 
single particle wave function uniquely.  We take the fitting of the 
angular distribution as an additional requirement, since it is  
sensitive to this wave function.
 
\begin{figure}[ht]
\epsfysize=110mm
\hspace*{3.0cm}
\epsffile{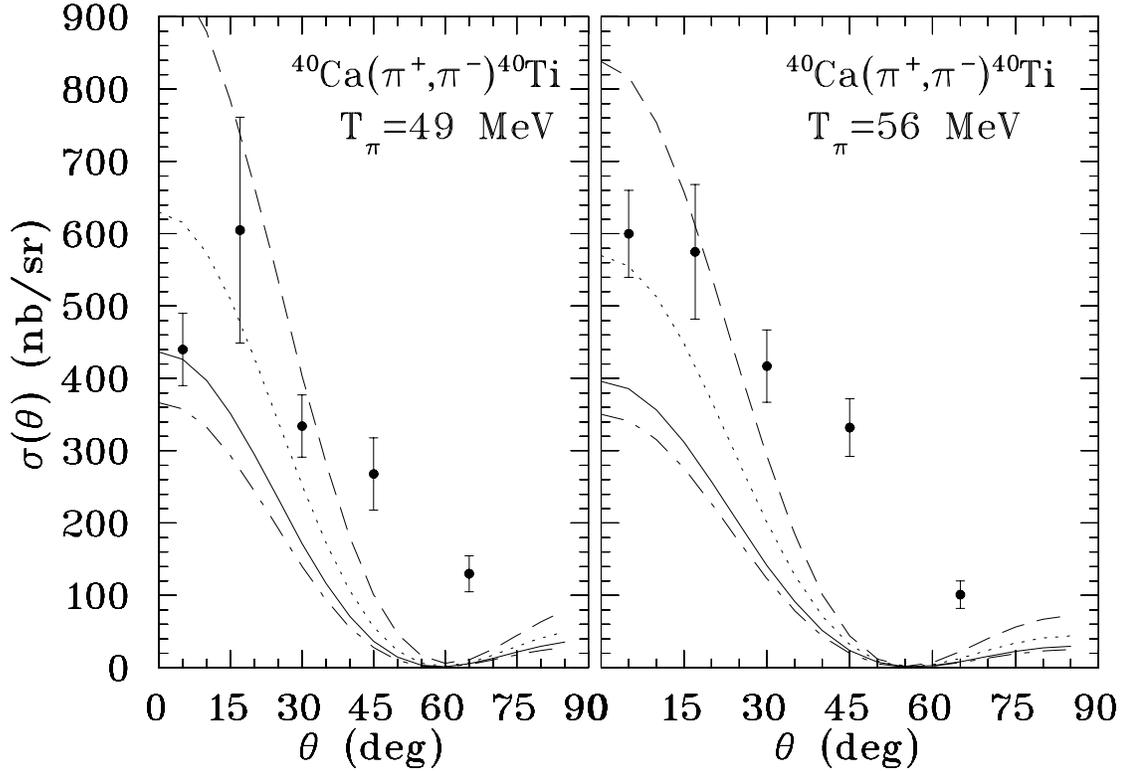}
\vspace*{-2.cm}
\caption{Angular distribution of \caftti at a pion laboratory energy of 49 
MeV and 56 MeV. The solid line represents the result of 
shell model, the dashed line gives the result of the SU(3) model, the 
dashed-dotted line the result of single particle shell model
and the dotted line the result of adding 5\% of \dfh\ neutron hole state.}
\label{f2}
\end{figure}

The angular distribution is calculated for \ostne at the pion laboratory
energy of 65 MeV (Fig. 1) and for the DCX reaction \caftti at energies of 
49 MeV and 56 MeV(Fig. 2). Data are taken from Ref. \cite{dra00}. For 
oxygen the agreement between theory and experiment is good except for 
the largest angle. For calcium the shape is approximately correct but
the magnitude is too small, especially at 56 MeV. As we shall see these
energies are on the sides of the peak so that slight changes in energy
change the magnitude a great deal.

\subsection{Energy Dependence}

Now that all the parameters are fixed, the energy dependence of the
forward amplitudes and cross sections can be calculated without any
adjustable parameters. As a first step, we calculate the forward
amplitude for all possible transitions between single particle orbits.  
A somewhat surprising feature is that there exist drastically different
patterns in some of the energy dependence of forward amplitudes.  For example,
Fig. 3 gives the energy dependence of the forward amplitudes of DCX on
$\ost$ for the pair transitions of $\poh \rightarrow \soh$ 
and $\poh \rightarrow \dth$, where the symbol $\poh \rightarrow
\soh$ means that a pair of neutrons in the \poh\ orbit is converted to
a pair of protons in the \soh\ orbit, etc. The dashed-dotted line, 
dashed line and solid line represent the real part, imaginary part 
and absolute value of amplitudes, respectively.
One can easily notice that the transition $\poh \rightarrow \soh$ has a
peak at low energy (around 50 MeV), whereas the transition $\poh
\rightarrow \dth$ has a peak at resonance.  
The origin of this drastic difference in pattern of the energy dependence
of forward amplitude is not yet understood, but one can expect from this
result that the energy dependence of the forward cross section will be
sensitive to the configuration mixing in the nuclear structure.

For DCX on $\caft$ \ the situation is somewhat different.  In Fig.  
\ref{f4} four DCX amplitudes are given:  \soh, \dth, \dfh\ and
\fsh$\rightarrow$\fsh.  While the first three transitions are involved in
the DCX on \caft, the fourth, \fsh$\rightarrow$\fsh\ is given only for
comparison. The first three transitions all have a very small value at
resonance region.  In contrast, the transition \fsh$\rightarrow$\fsh\ has
a second peak at resonance.

The forward cross sections of DCX on \ost \ and \caft \ are calculated
with the configuration mixing ratio provided by the three nuclear models
in Table 1. From table one can easily see the relevant transitions
which are: \poh, \pth $\rightarrow$ \soh, \dth\ and \dfh\ for \ost, and
\soh, \dth\ and \dfh $\rightarrow$\fsh \ for \caft, respectively. The
results of DCX forward scattering are given in Figures \ref{f5} and
\ref{f6}. The solid line, dashed line and dashed-dotted line represent
results of the shell model, the SU(3) model and the single particle shell
model, respectively.

\begin{figure}[ht]
\epsfysize=120mm
\vspace*{-1.cm}
\hspace*{1.5cm}
\epsffile{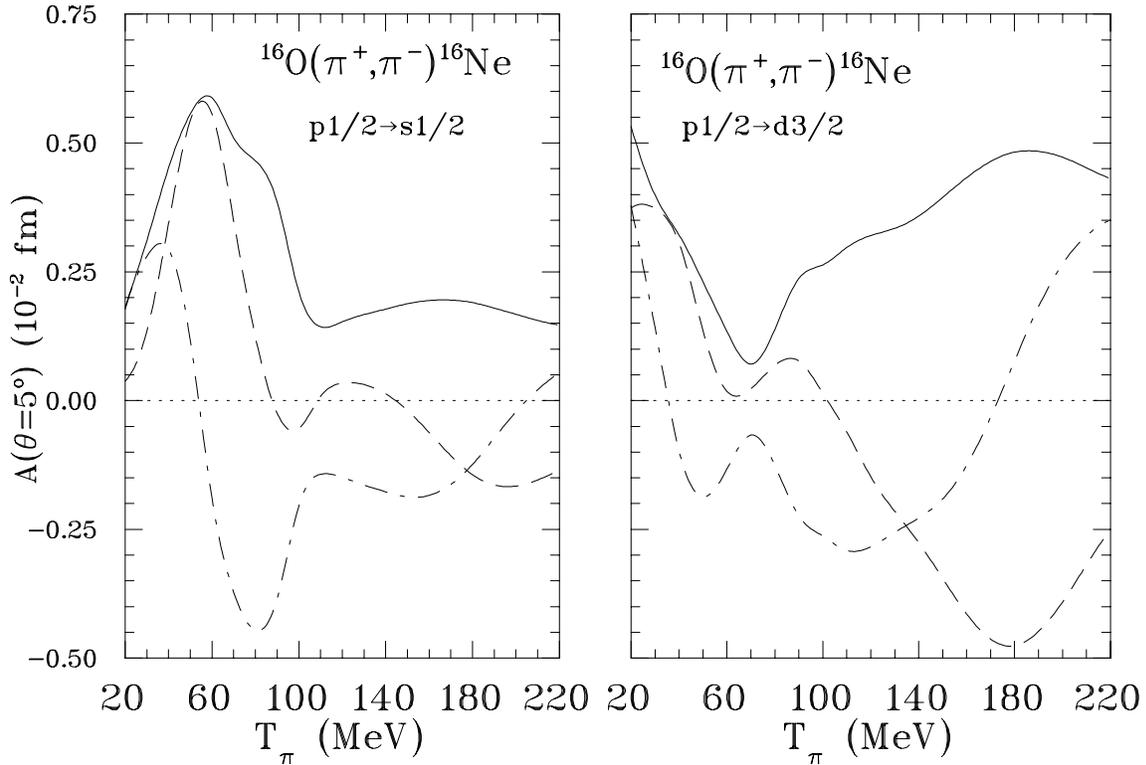}
\vspace*{-1.cm}
\caption{Examples of the energy dependence of forward 
amplitudes for DCX on \ost. Transitions are from the  \poh \ orbit to ds
orbits. The dashed-dotted, dashed and solid lines represent the real part, 
the imaginary part and the absolute value, respectively. 
Optical model parameters are taken from fits to the elastic scattering.
}
\label{f3}
\end{figure}

\begin{figure}[ht]
\epsfysize=120mm
\vspace*{-1.cm}
\hspace*{1.5cm}
\epsffile{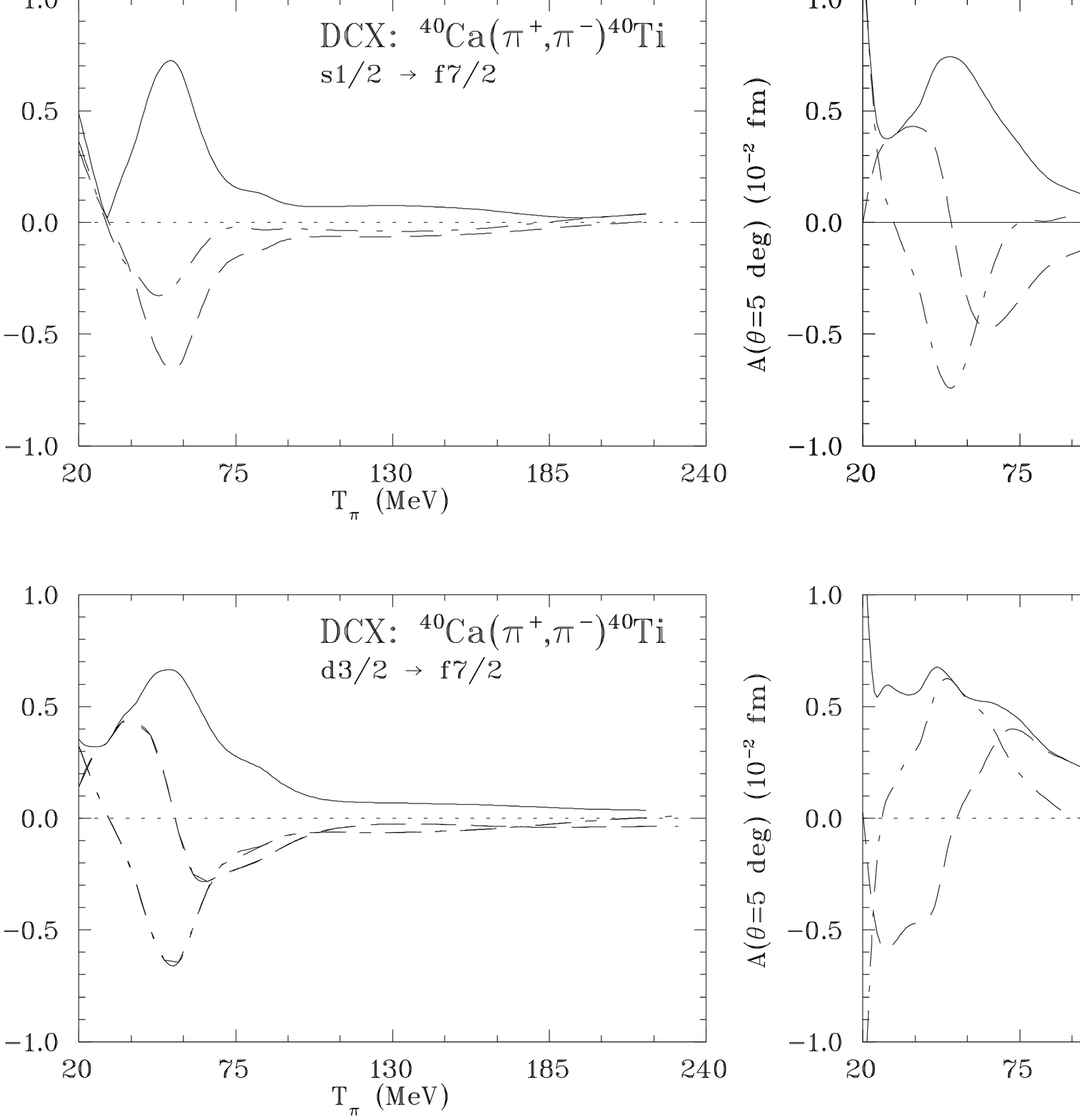}
\vspace*{-1.cm}
\caption{Energy dependence of forward 
amplitudes for DCX on \caft. Transitions are from  ds orbits to the \fsh\  
orbit. The dashed-dotted, dashed and solid lines represent the real part, 
the imaginary part and the absolute value, respectively. 
Optical model parameters are taken from fits to the elastic scattering.
}
\label{f4}
\end{figure}

\begin{figure}[ht]
\epsfysize=120mm
\hspace*{2.0cm}
\epsffile{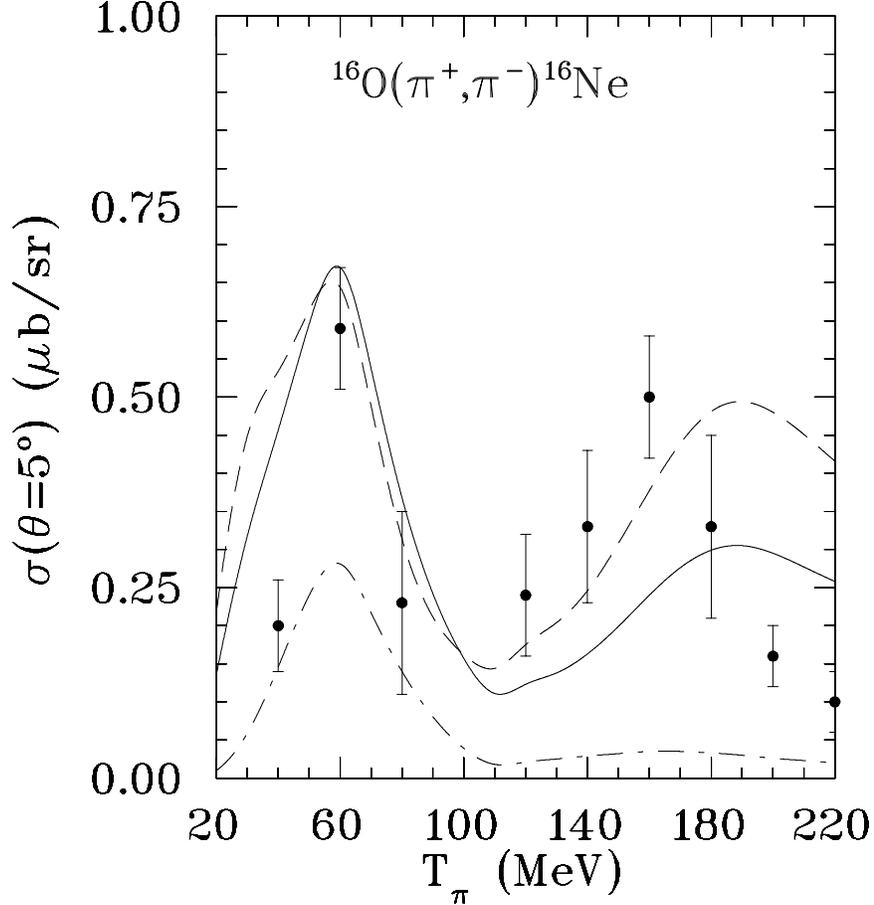}
\caption{Energy dependence of forward DCX cross section on $\ost$.  
For the meaning of the curves see the caption of Fig.1.}

\label{f5}
\end{figure}

\begin{figure}[ht]
\epsfysize=120mm
\hspace*{2.0cm}
\epsffile{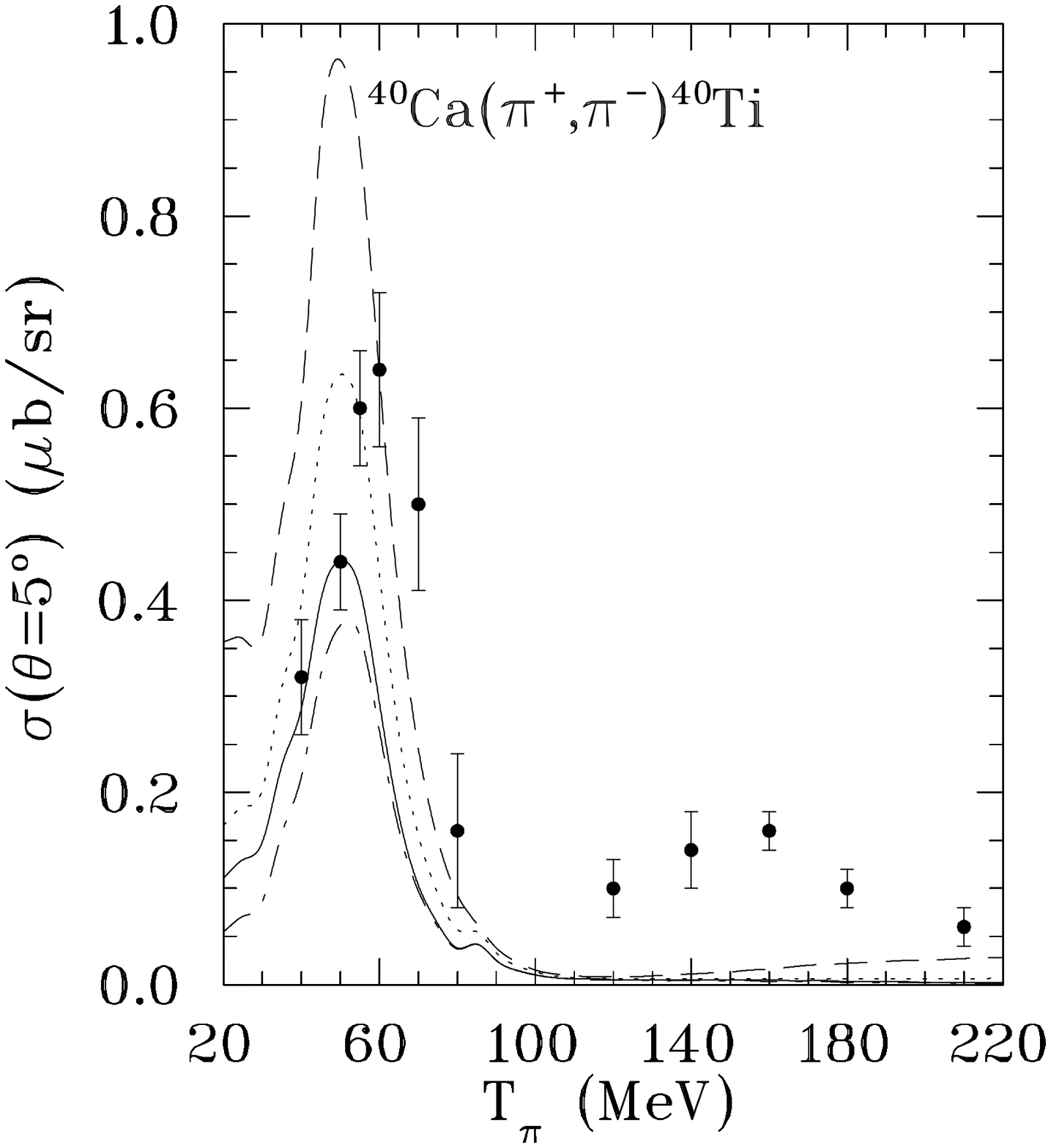}
\caption{Energy dependence of forward DCX cross section on $\caft$.  
See the caption of Fig. \ref{f2} for the meaning of the curves.}
\label{f6}
\end{figure}

Figure \ref{f5} shows that with the shell model wave function (solid line)  
both peaks at low energy and at resonance are reproduced fairly well.  While
the SU(3) result (dashed line) is close to that of shell model, the single
particle shell model gives a poor prediction.  If one believes the optical
model we use, then the energy dependence of the forward DCX reaction suggests
that the SU(3) model is a good approximation for the wave function of
$\nest$.

For DCX on \caft, Fig. \ref{f6} demonstrates that the peak at low energy
is reproduced by all three nuclear models but with different height.  
While the peak of shell model is lower than the experimental one, that of
SU(3) is higher than the data.  This suggests that our shell model
calculation needs to be improved.  A possibility that suggests itself is
the addition of the configuration mixing of $(\dfh)^{-2}$. A mixing of
this kind with 5\% probability was tried, and the calculation gives a
peak at low energy much closer to experiment. The angular distributions
are also improved considerably. The improved results for angular
distribution and forward cross section are shown with a dotted line in
figures \ref{f2} and \ref{f5}, respectively.  This mixing of $(\dfh)^{-2}$
seems reasonable, since its neglect in subsection 2.1 was only because of
the absence the single hole state $(\dfh)^{-1}$ in the spectrum of
$^{39}$Ca. However, due to pairing between neutrons, the ground state of
$^{38}$Ca may well contain a significant mixing of the component of
$|(\dfh )^{-2};0>$. However, the adding of the configuration \dfh \ for
\tift \ needs to be justified by independent nuclear structure studies.  
Here we take this as an example of the sensitivity of DCX amplitudes to
configuration mixing.

In both figures \ref{f5} and \ref{f6}, although the height varies, the
peaks appear at low energy for all nuclear models, which is consistent
with previous claims that this peak is the consequence of the pion wave
distortion.

While the position and height of the peak around 50 MeV are reproduced
reasonably well, there are two discrepancies concerning the second peak, i.e.
the peak at around resonance.  First the position of the calculated second
peak for \ost \ appears at around 180 MeV whereas the experimental one is at
around 160 MeV.  In a two-step sequential process (which is the framework of
this paper)  the second peak is produced by the resonant behavior of the
p-wave SCX amplitude which corresponds to the delta-resonance although the
explicit degree of freedom of the delta is not included. To shift the peak
from 180 to 160 MeV one may need to explore the processes beyond the degrees
of freedom of nucleons.

The second discrepancy is that the calculation of\ \caft \ shows almost no
peak at resonance. Among the three transitions, \soh, \dth, and
\dfh$\rightarrow$\fsh, which are involved in the DCX on \caft, the first
two give almost no peak at the resonance (see Fig. \ref{f4}), and the
third one has a scarcely noticeable DCX amplitude in this region.  This
means that it is impossible to get a good fit in the resonance region by
only adjusting the nuclear wave functions, in contrast to the case of
\ost, for which the magnitude of the DCX amplitudes at resonance is
comparable to the low energy region.  To reproduce the second peak on
\caft\ we may have to develop a better understanding of the energy
dependence of DCX amplitude for transitions between orbits.

Another possible remedy is to readjust the true absorption, since one of the
consequence of true absorption is to reduce the amplitude at resonance.  
In this study all the optical model parameters, i.e. the ranges of six
partial waves, the absorption and the s and p wave multipliers are taken
without modification from the corresponding elastic $\pi$-nucleus
scattering fit \cite{nus98a}. The readjustment of the true absorption
requires a re-calculation of the elastic $\pi$-nucleus scattering.

Despite all these problems, we can say that the fair agreement between 
data and theoretical calculation, at least for low energies,  
is an indication of the validity of the optical model employed. 

\section{Double Isobaric Analog State Transition}

The DCX reaction \oetne has been investigated experimentally by a number 
of authors \cite{gre82,for89} with an emphasis on pion energies above 100 
MeV, leaving the peak at low energy poorly explored.

The structure of \oet \ and \neet \ has been studied based on the analysis of
spectra and transitions \cite{law76,for89}, and it has been shown that the
ground states of \oet \ and \neet \ contain a significant mixing between 
two particle and four-particle-two-hole configurations. The ground state 
of \oet\ can be written as \cite{law76},
\eq
|0^{+}_{1}>=0.842|(n{\dfh})^2 >+0.440|(n \soh)^2> -0.313|\Psi_{00}>, 
\qe
where the $|\Psi_{00}>$ is a deformed state formed by exciting a pair of
protons from the \poh \ shell to the ds shell. 
The $|\Psi_{00}>$ wave function was given in Ref. \cite{law76} 
as a mixture of different four-particle-two-hole states. 
For the sake of simplicity, we consider only the angular momentum 
zero pairs while keeping the total wave function  $|\Psi_{00}>$
normalized. Therefore, in this approximation,
\eqa
  |\Psi_{00} > &=& \{ \frac{4}{15} |(p \dfh)^2 (n \dfh)^2 >
+ \frac{8}{45} |(p \dth)^2 (n \dth)^2 > 
+ \frac{5}{9} |(p \soh)^2 (n \soh)^2 > \nonumber \\
 & & + \frac{8}{15\sqrt{3}}|(p\dfh)^2 (n\dth)^2+(n\dfh)^2 (p\dth)^2> 
 \nonumber \\
 & &+ \frac{2\sqrt{2}}{3\sqrt{3}}|(p\dfh)^2 (n\soh)^2+(n\dfh)^2 
(p\soh)^2>\nonumber \\
& &+ \frac{4}{9}|(p\dth)^2 (n\soh)^2+(n\dth)^2 (p\soh)^2> \} 
|(p\poh)^{-2}>,
\qea
where the superscript 2 (--2) represents a two-particle (two-hole) state, 
and the angular momenta of all the pairs are taken to be zero.
The ground state of \neet \ is assumed to be the same as \oet \ except
that neutrons are changed to protons and vice versa.

We take the optical model parameters from \ost \ \cite{nus98a}. 
As was done in the case of the GS transition on \ost \ and \caft \ we fit the
angular distribution in order to determine the single particle wave
function. In the case of the DIAS transition on \oet \ one has to pay more
attention to the single particle wave functions for the following reasons.

The structure of \oet \ is quite similar to that of \caftt: two neutrons
outside of the doubly closed shell. In a rough approximation both nuclei
can be described as a two-particle state. Nuclear structure studies 
\cite{law76,flo68} have shown that they both have a mixing between two 
particle states and four-particle-two-hole states, even the mixing rates
are quite similar one to another. It has been found for \caftt \
\cite{gib92} that the radius of the valence neutrons is much larger than
what one would expect from a calculation with a Woods-Saxon type of
potential. To be more specific, a Woods-Saxon potential calculation for
\caftt \ gives an rms radius of the \fsh\ valence neutron of 3.95 fm,
whereas the analysis in Ref. \cite{gib92} of pion scattering data gives a
value for this radius of 4.95 fm.  This one femtometer difference may be
understood as a consequence of the fact that the two neutrons form a
kind of cluster, which increases the attraction between the two valence
neutrons and diminishes the interaction between the nucleon pair and the
doubly closed core.  Due to the similarity in nuclear structure between
\oet \ and \caftt \ one may suspect that the same phenomenon occurs in
\oet, i.e. an rms radius of valence neutrons considerably larger than that
derived from a Woods-Saxon potential.

\begin{figure}[ht]
\epsfysize=130mm
\vspace*{-.5cm}
\epsffile{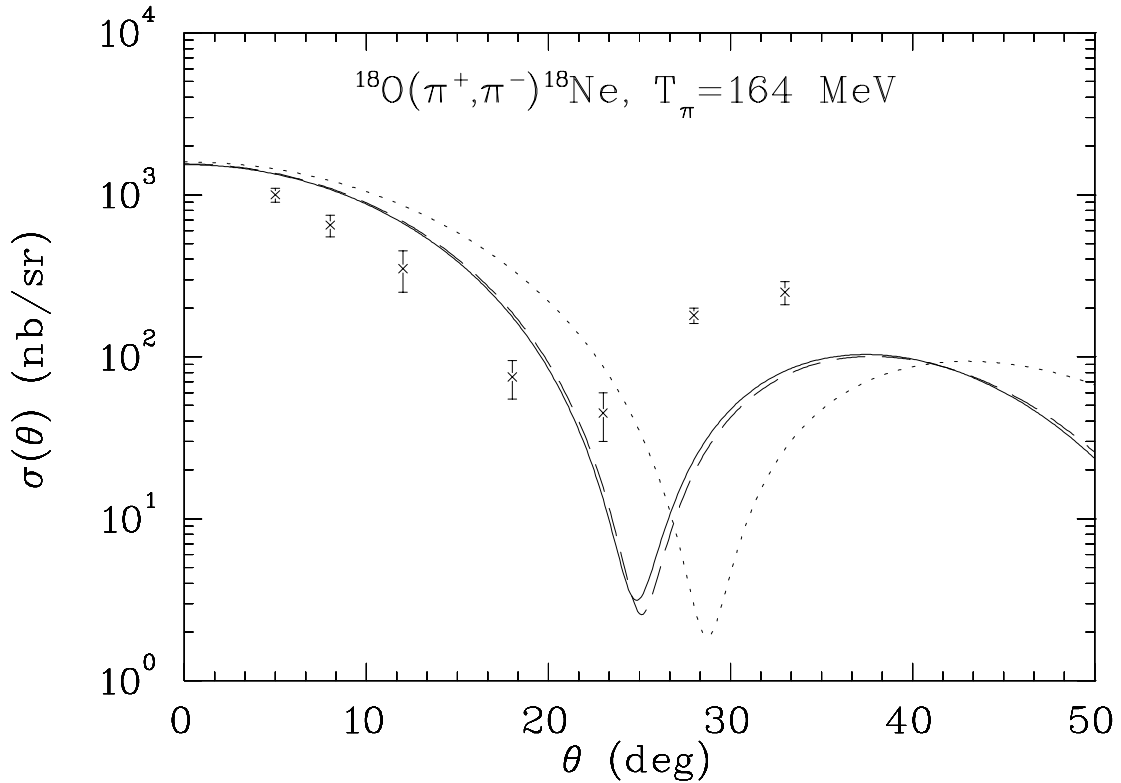}
\vspace*{-2.cm}
\caption{Angular distribution of the DCX of \oetne at the pion laboratory
energy of 164 MeV. The solid line presents the results with a WS1 type of
wave function, the dotted line shows the WS type of wave function and the
dashed line corresponds to the HO1 type of wave function. See the text for 
the definition of types of wave functions}
\label{f7}  
\end{figure}

\begin{figure}[ht]
\epsfysize=120mm
\hspace*{2.0cm}
\epsffile{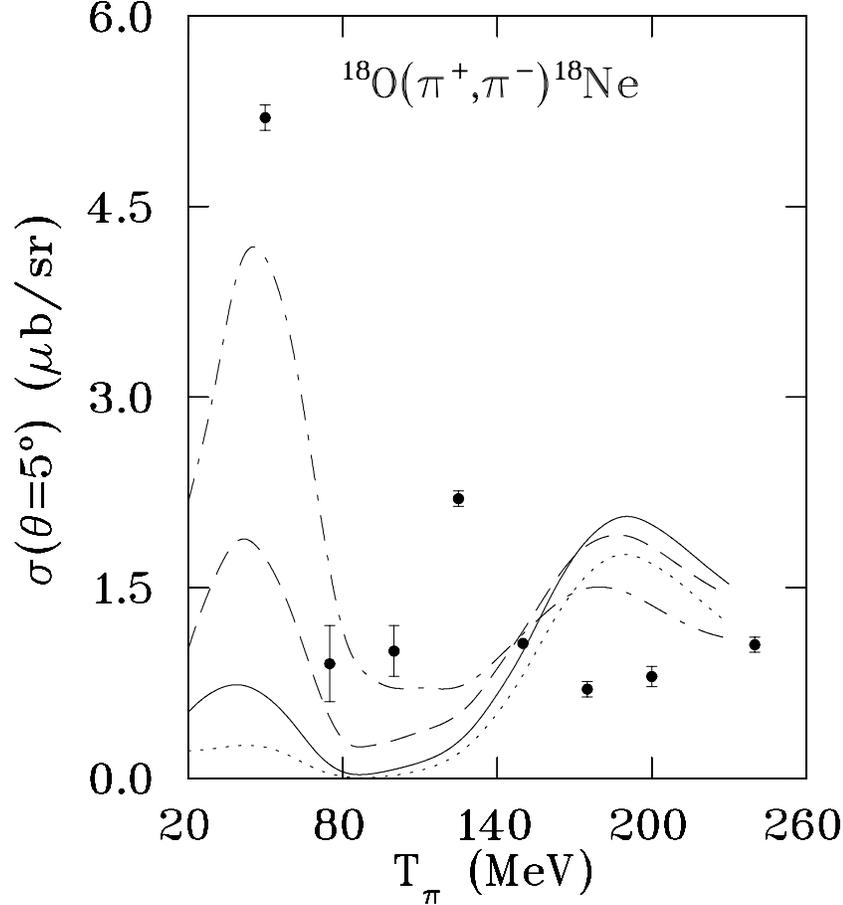}
\vspace*{-.3cm}
\caption{Energy dependence of forward DCX cross section on $\oet$.
The dashed-dotted line, the dashed line, the solid line, and the
dotted line correspond to the HO0,HO, HO1 and HO2 type of wave
functions, respectively. See text for the definition of the wave
function type.} 
\label{f8}
\end{figure}   

With this assumption in mind, the angular distribution at pion energy 
164 MeV is calculated with three types of single particle wave functions.
The first type is from a standard calculation with a Woods-Saxon type of
potential as described in Section 2. The rms radii of the 1d orbit
($r_{1d}$), the 2s orbit ($r_{1d}$) and the total neutron sector ($r_n$)
are 3.68 fm, 3.91 fm and 2.86 fm, respectively. This wave function is
denoted as the WS wave function, and the angular distribution calculated
with this wave function is given in Fig. \ref{f7} by the dotted line.  
The second type is made by combining the \ost \ core of the WS wave
function and a valence wave function from an expanded Woods-Saxon
potential well.  Its radii are given in Table 2. This type is denoted as
WS1 and the resulting cross section is shown in Fig. \ref{f7} by the
solid line. A third wave function is made in a similar way as the WS1
type except that the valence wave function is from a harmonic oscillator
potential well. Its radii are given in Table 2 under the label HO1 and
the corresponding angular distribution is given in Fig. \ref{f7} by the
dashed line. The radii of the WS1 and HO1 wave functions are chosen to be
very close in order to explore the dependence of angular distribution on
details of wave function with their radii being the same. We note that in
all cases only the shell model nuclear wave function (described by Eqs.
(11) and (12)) was used.  The experimental data are taken from Ref.
\cite{gre82}.

The WS and WS1 types of wave functions differ in their $r_{1d}$ and
$r_{2s}$ radii. A comparison between them shows that with increasing
radius the angle of the first minimum decreases, and the agreement with
data is improved considerably.  This can be viewed as a strong support for
the assumption of the enlargement of the radius of the valence neutron.
Radii of $r_{1d}=4.66$\ fm and $r_{2s}=4.42$\ fm would appear to be  
reasonable estimates for the valence nucleons.

A comparison of the cases of HO1 and WS1 shows that these two wave
functions, with almost equal radii for valence neutron, give nearly the
same angular distribution despite the differences in the shape of their 
wave functions. Therefore, one may expect that in the study of the DCX 
reaction (at least at this energy) the nuclear radius plays a crucial 
role while the details of the wave function are of little importance.

To further explore the effects of the enlargement of valence neutron
radius we calculated the energy dependence of the forward DCX with the
following four types of wave function: HO0, HO, HO1 and HO2,
where the HO1 type was explained above and the HO0, HO and HO2 are made in
the same way as of the type of HO1 but with different radii. The cross
sections resulting from wave functions of the types HO0, HO, HO1 and HO2
are represented in Fig. 7 by dashed-dotted, dashed, solid and dotted line,
respectively. We list the radii for all types of wave function in Table 2.  
We used the harmonic oscillator type of wave function for the valence
neutron for convenience, and it should not give results significantly
different from those with an expanded Woods-Saxon potential.

A very interesting feature of Fig. \ref{f8} which shows the forward angle
energy dependence is that the height of the peak at 50 MeV increases with
a decreasing radius of valence neutron while the height of the second peak
at resonance is little changed. The following could be a possible
explanation:  The angular correlation is being held fixed (i.e.  the shell
model wave function) so, as the radius is increased the two nucleons are
getting farther apart.  The DCX cross section drops rapidly as that
happens (like the inverse of the distance between them squared). Thus, at
low energy where the absorption effects are less important, the cross
section rises as the radius is decreased. On the other hand, as the radius
is increased both nucleons are farther from the nucleus and the
probability of being absorbed is decreased. This effect is very strong in
the resonance region so that around the delta the two effects compete and
tend to cancel.

The forward scattering shown in Fig. \ref{f8} seems to favor a smaller
radii which would contradict what was suggested by the angular
distribution.  However, the above treatment is only schematic, a more
sophisticated self-consistent calculation is needed to understand the
enlargement of the neutron radius of \oet.

\begin{table}[ht]
\vspace*{-12pt}
\begin{center}
\caption[]{Radii of the valence neutron and the total neutron sector for 
\oet. See the text for the meaning of the wave function type.}
\vspace*{14pt}
\begin{tabular}{|c|c|c|c|}
\hline
wave function type & $r_{1d}$ & $r_{2s}$ & $r_{n}$ \\ \hline
\hline
 WS &  3.68 & 3.91 & 2.86 \\
 WS1& 4.66 & 4.42  & 3.14 \\ \hline
 HO0 & 3.22 & 3.22 & 2.75 \\
 HO &  3.70 & 3.70 & 2.87 \\
 HO1 & 4.65 & 4.65 & 3.13\\
 HO2 & 5.69 & 5.69 & 3.46 \\ \hline
 \end{tabular}
\label{table}
\end{center}   
\end{table}

The calculation reproduces only the gross feature of experimental data.  
As in the case of the GS transition, a peak appears at low energy (around
50 MeV), but the height of the peak is much lower than the experiment
if one takes the angular-distribution favored radii ($r_{1d}=4.66$ and 
$r_{2s}=4.42$ fm). Another discrepancy is that the second peak is 
calculated to be around resonance, but the experimental one is around 120 
MeV.

In Ref. \cite{gre82} an attempt was made to reproduce the peak at 120 MeV
with a two-amplitude model, which mixes the DIAS amplitude of \oet \ and
the GS transition amplitude of \ost.  Although it was only a
schematic treatment, the reproduction of the second peak is an indication
that a configuration mixing more comprehensive than what was done in this
work might be successful.

\section{Conclusion and Discussion}

In summary, the present work has demonstrated that the conventional
sequential process can reproduce fairly well the angular distribution and
the energy dependence of forward DCX reactions on both the GS transition
of \ost \ and \caft \ and the DIAS transition on \oet.  In our
calculation the peak at low energy arises naturally as the consequence of 
pion wave distortion without invoking explicit dibaryon degrees of freedom, 

It is often thought that the DCX reaction depends mainly on the reaction
mechanism with detailed nuclear structure being unimportant. However, this
work shows that nuclear structure is also important in describing the DCX
reaction.  Due to a drastic difference in the pattern of energy dependence of
forward amplitudes among different transitions between single particle
orbits, the DCX reaction is sensitive to the configuration mixing ratios in
the nuclear structure. In this work three nuclear models were employed.  
While all three models predict a peak at around 50 MeV, the height of the
peak is quite different among them. The shell model gives fairly good
predictions on the forward DCX reaction both for $\ost$ and $\caft$.  The
difference among the results of different nuclear models can give an
indication of how large a role the nuclear structure plays in the DCX
reactions.  In other words, the energy dependence of the forward DCX reaction
contains information not only on the fundamental charge exchange process but
on the nuclear structure as well.

In this work only the nucleon pairs with angular momentum zero were
considered, i.e., only the elementary process $(nn)^{(0)}
(\pi^{+},\pi^{-}) (pp)^{(0)}$ was involved, where the superscript 0
represents the angular momentum of the nucleon pair. We also made some
calculations including the pairs with angular momenta 2 and 4, but no
significant change in the DCX amplitude was found.  However, the role of
higher angular momentum pairs can not be excluded definitively.

We looked at the consequences of enlarging the radii of orbitals in \oet\ 
as might be expected from pairing of the valence neutrons. We found that
such considerations can solve a significant fraction of the problem of the
position of the first minimum but give worse agreement with the energy
dependence of the forward cross section.

The study in the present work can be extended to medium mass nuclei, as
data show that the forward cross section of the DCX reactions $^{56}{\rm
Fe}(\pi^{+},\pi^{-})^{56}{\rm Ni}\ $ and $^{93}{\rm
Nb}(\pi^{+},\pi^{-})^{93}{\rm Tc}\ $ also has a peak around 50 MeV. These
nuclei have a much larger shell model space which makes an exact shell
model calculation almost impossible. However, some recent developments in
nuclear structure theory can reduce the space substantially, which
makes the study possible.  One of the new models is the pseudo-SU(4)  
model \cite{isa99} which deals with nuclei with nucleons in the pf-shell
by using the technique of the SU(4) model while incorporating the large
L-S coupling.  We intend to extend the model to the mass region $28 \leq
Z,N \leq 50$, and a study of the application of this model to DCX in
this mass region is currently underway.

\section*{Acknowledgments}

The work is supported by the University of Antioquia, Colombia, and by  the 
National Science Foundation under contract PHY-0099729.

\end{document}